\documentclass{article}
\usepackage{amssymb}

 \def\rql{\mbox{\strut\smash{\raisebox{-1.4ex}{''}}\hspace{-0.05em}}}
 \def\rqr{\mbox{\hspace{0em}``}}

\newcommand*{\mytheorem}{\mbox{T~h~e~o~r~e~m.~}}

\makeatletter
\renewcommand{\@oddfoot}{--- ~March, 2006~ ---\hfil
\raisebox{0.3ex}{\tiny --- ~D.~A.~Arbatsky~ ``TCP II: Proof of the
uncertainty principle''~
---}\hfil --- ~\thepage~ ---}
 \makeatother

\begin{document}


\title{The certainty principle II\\
Proof of the uncertainty principle}
\author{D.~A.~Arbatsky\footnote{ http://daarb.narod.ru/ , http://wave.front.ru/ }}
\date{March, 2006}
\maketitle

\begin{abstract}
A more detailed derivation of the Heisenberg uncertainty principle
from the certainty principle is given.
\end{abstract}


\paragraph{Introductory remarks.}
After publication of the paper~\cite{Arb2005tcp} I received many
letters, including those with references. In this connection, I
think it is necessary to specify the following:
\begin{enumerate}
\item
The metric introduced by me (``quantum angle'') is known to
mathematicians from 1904 as {\it Fubini-Study metric}.
\item
The correct ``uncertainty relation'' (in fact, {\it certainty}
relation) for the quantities energy - time was first suggested by
Mandelshtam and Tamm~\cite{MandTamm1945}.
\item
Mandelshtam and Tamm studied a quantum system in the Schr\"odinger
representation and did not use group theory methods. For this
reason they could not understand that their result has more
general character.
\item
In contrast, I used group theory formulations. And implied that
the system can be studied, in particular, in the representation of
relativistic canonical quantization~\cite{Arb2005wircq}. This
allowed me to formulate the certainty principle and to suggest
more general inequalities.
\end{enumerate}

Some of my critics refused to believe that the Heisenberg
uncertainty principle is really a consequence of the certainty
principle. They claimed that ``this can not be so, because this
can never be so''.

Nevertheless, this is so. And here I give a more detailed
explanation.


\paragraph{Derivation of the uncertainty principle.}
Suppose that for a given quantum system we succeeded to find some
observable $X$, that can be considered in some sense a
``coordinate operator''.

Suppose that $X$ is a self-adjoint operator with continuous
spectrum, $X=X^*$. Let us denote $\Omega_{(a,b)}$ its spectral
projector\footnote{Roughly speaking, spectral projector is an
operator nullifying wave function in $X$-representation outside of
the given interval.} for an arbitrary real interval $(a,b)$.

Suppose also that we have a self-adjoint operator $P$, $P=P^*$,
such that for any $a$, $b$ and $\delta x$ we have equality:
\[
 e^{+\,i\,\delta x\,P/\hbar} \,
  \Omega_{(a+\delta x,b+\delta x)} \,
    e^{-\,i\,\delta x\,P/\hbar} \,=\,
 \Omega_{(a,b)} \ ,
\]
i.~e. $P$ is a ``generator of spectral shifts'' for $X$. As we
know, such an operator is usually an operator of momentum.

Suppose now that the system is in state $\rangle$,
$\langle\,|\,\rangle=1$.

The quantity $\langle\,\Omega_{(a,b)}\,\rangle$, obviously,
defines the probability to find the system inside the interval
$(a,b)$. Let us define such $l$ and $r$, that
\[
 \textstyle
 \langle\,\Omega_{(-\infty,l)}\,\rangle
 \,=\,
 \langle\,\Omega_{(r,+\infty)}\,\rangle
 \,=\,
 \frac12-\frac12\sqrt{1-\cos^21} \,\approx\,
 0,07926\dots
\]
It is easy to see, that $l$ and $r$ exist\footnote{But, generally
speaking, they are not unique. In order to eliminate this
non-uniqueness, it is convenient to choose $l$ maximum of the
possible, and $r$~--- minimum. Then the distance $r-l$ will be
minimum.}. So, the quantity $\delta_\rangle X = r-l$ can be
naturally called ``uncertainty'' of the coordinate $X$.

\mytheorem{}
 {\it The following inequality takes place (the uncertainty principle):
\begin{equation}
 \begin{array}{|ccc|}
 \hline
 \phantom{1} & & \phantom{1}\\
  & \delta_\rangle X \Delta_\rangle P \geqslant \hbar & \\
 \phantom{1} & & \phantom{1}\\
 \hline
 \end{array}
 \label{Certainty}
\end{equation}}

In order to prove this theorem let us first estimate the scalar
product of the vector $\rangle$ and the shifted vector
$e^{-\,i\,\delta_\rangle X\,P/\hbar} \,\rangle$:
\[
 \big|\,
  \langle\,
   |\,
   e^{-\,i\,\delta_\rangle X\,P/\hbar}
  \,\rangle
 \,\big|
 \,=\,
 \big|\,
  \langle\,
   (\,\Omega_{(-\infty,r)} + \Omega_{(r,+\infty)}\,)
   \,
   e^{-\,i\,\delta_\rangle X\,P/\hbar}
  \,\rangle
 \,\big|
 \,=\,
\]
\[
 \,=\,
 \big|\,
  \langle\,
   \,\Omega_{(-\infty,r)}\, e^{-\,i\,\delta_\rangle X\,P/\hbar}
  \,\rangle
   \,+\,
  \langle\,
   \Omega_{(r,+\infty)}\, e^{-\,i\,\delta_\rangle X\,P/\hbar}
  \,\rangle
 \,\big|
 \,\leqslant\,
\]
\[
 \,\leqslant\,
 \big|\,
  \langle\,
   \,\Omega_{(-\infty,r)}\, e^{-\,i\,\delta_\rangle X\,P/\hbar}
  \,\rangle
 \,\big|
   \,+\,
 \big|\,
  \langle\,
   \Omega_{(r,+\infty)}\, e^{-\,i\,\delta_\rangle X\,P/\hbar}
  \,\rangle
 \,\big|
 \,=\,
\]
\[
 \,=\,
 \big|\,
  \langle\,
   \,(\Omega_{(-\infty,r)})^2\, e^{-\,i\,\delta_\rangle X\,P/\hbar}
  \,\rangle
 \,\big|
   \,+\,
 \big|\,
  \langle\,
   (\Omega_{(r,+\infty)})^2\, e^{-\,i\,\delta_\rangle X\,P/\hbar}
  \,\rangle
 \,\big|
 \,=\,
\]
\[
 \,=\,
 \big|\,
  \langle\,
   \,\Omega_{(-\infty,r)}\,
   e^{-\,i\,\delta_\rangle X\,P/\hbar}
   \,\Omega_{(-\infty,l)}
  \,\rangle
 \,\big|
   \,+\,
 \big|\,
  \langle\,
   \Omega_{(r,+\infty)}\,
   e^{-\,i\,\delta_\rangle X\,P/\hbar}
   \,\Omega_{(l,+\infty)}
  \,\rangle
 \,\big|
 \,\leqslant\,
\]
\[
 \,\leqslant\,
  \langle\,
   \Omega_{(-\infty,r)}
  \,\rangle^{1/2}\,
  \langle\,
   \Omega_{(-\infty,l)}
  \,\rangle^{1/2}\,
   \,+\,
  \langle\,
   \Omega_{(r,+\infty)}
  \,\rangle^{1/2}\,
  \langle\,
   \Omega_{(l,+\infty)}
  \,\rangle^{1/2}\,
 \,=\,
\]
\[
 \textstyle
 \,=\,
 \sqrt{\frac12+\frac12\sqrt{1-\cos^21}}
 \cdot
 \sqrt{\frac12-\frac12\sqrt{1-\cos^21}}
 \ + \qquad\qquad\qquad\qquad
\]
\[
 \textstyle
 \qquad\qquad\qquad\qquad + \,
 \sqrt{\frac12-\frac12\sqrt{1-\cos^21}}
 \cdot
 \sqrt{\frac12+\frac12\sqrt{1-\cos^21}}
 \,=\,
 \cos 1
 \ .
\]
Here, coming from the fifth to the sixth line, we estimated both
terms by the Cauchy-Bunyakovsky-Schwarz inequality.

So, for the quantum angle between the initial and the shifted
vectors we have estimation:
\[
 \angle\,\bigl(\
  \rangle
  \ , \
  e^{-\,i\,\delta_\rangle X\,P/\hbar}\,\rangle
 \ \bigr)
 \ \geqslant\ 1 \ .
\]
But it means that under the action of
 $e^{-\,i\,\delta_\rangle X\,P/\hbar}$
the vector $\rangle$ changes {\it
substantially}~\cite{Arb2005tcp}.

Applying the {\it certainty}\/ principle~\cite{Arb2005tcp}, we
directly get~(\ref{Certainty}). $\blacksquare$


\paragraph{Discussion.}
Historically, it stacked up so~\cite{HilgUff2001} that now in all
textbooks the Heisenberg uncertainty principle is illustrated with
the help of the Kennard inequality,
\[
 \Delta_\rangle X\,\Delta_\rangle P \,
 \geqslant\, \frac\hbar2 \ \ ,
\]
which is not identical with~(\ref{Certainty}). Its undoubted
virtue is that its proof is easier to understand for a person, who
just starts to study quantum mechanics. Nevertheless, I believe
that~(\ref{Certainty}) is more fundamental.


\paragraph{Acknowledgements.}
In closing I want to thank T.~A.~Bolokhov, A.~V.~Ossipov,
E.~V.~Aksyonova, A.~Yu.~Tos\-che\-vi\-ko\-va, A.~Kleyn, M.~Gatti,
A.~K.~Pati, and P.~Enders for helpful discussions, references and
help in procuring literature.



\end{document}



\title{Принцип определённости II\\
Доказательство принципа неопределённости}
\author{Д.~А.~Арбатский\footnote{ http://daarb.narod.ru/ , http://wave.front.ru/ }}
\date{Март 2006 г.}
\maketitle

\begin{abstract}
Приводится более детальный вывод принципа неопределённости
Гейзенберга из принципа определённости.
\end{abstract}


\paragraph{Вводные замечания.}
После опубликования статьи~\cite{Arb2005tcp} я получил много
отзывов, в том числе с литературными указаниями. В связи с этим,
считаю необходимым уточнить следующее:
\begin{enumerate}
\item
Введённая мною метрика (\rql квантовый угол\rqr) известна
математикам с 1904 г. как {\it метрика Фубини-Штуди}.
\item
Правильное \rql соотношение неопределённостей\rqr{} (на самом
деле, соотношение {\it определённости}) для величин энергия -
время было впервые предложено Мандельштамом и
Таммом~\cite{MandTamm1945}.
\item
Мандельштам и Тамм изучали квантовую систему в представлении
Шрёдингера, а также не использовали теоретико-групповые методы.
Поэтому им не удалось понять, что их результат имеет более общий
характер.
\item
Я же использовал теоретико-групповые формулировки. И подразумевал,
что система может, в частности, изучаться в представлении
релятивистского канонического квантования~\cite{Arb2005wircq}. Это
позволило мне сформулировать принцип определённости и предложить
более общие неравенства.
\end{enumerate}

Некоторые мои критики отказались верить, что принцип
неопределённости Гейзенберга действительно является следствием
принципа определённости. Они заявили, что \rql этого не может
быть, потому что этого не может быть никогда\rqr.

Тем не менее, это так и есть. И здесь я даю более подробное
объяснение.


\paragraph{Вывод принципа неопределённости.}
Допустим, у рассматриваемой квантовой системы нам удалось найти
некоторую наблюдаемую $X$, которую можно в том или ином смысле
считать её \rql оператором ко\-ор\-ди\-на\-ты\rqr.

Будем полагать, что $X$~--- самосопряжённый оператор с непрерывным
спектром, $X=X^*$. Обозначим $\Omega_{(a,b)}$ его спектральный
проектор\footnote{Грубо говоря, спектральный проектор~--- это
оператор, зануляющий волновую функцию в $X$-представлении вне
данного интервала.} для произвольного вещественного интервала
$(a,b)$.

Допустим также, что имеется самосопряжённый оператор $P$, $P=P^*$,
такой, что для любых $a$, $b$ и $\delta x$ имеет место равенство:
\[
 e^{+\,i\,\delta x\,P/\hbar} \,
  \Omega_{(a+\delta x,b+\delta x)} \,
    e^{-\,i\,\delta x\,P/\hbar} \,=\,
 \Omega_{(a,b)} \ ,
\]
т.~е. $P$ является \rql генератором спектральных сдвигов\rqr{} для
$X$. Как известно, таким оператором обычно является оператор
импульса.

Пусть теперь система находится в состоянии $\rangle$,
$\langle\,|\,\rangle=1$.

Величина $\langle\,\Omega_{(a,b)}\,\rangle$, очевидно, определяет
вероятность найти систему внутри интервала $(a,b)$. Определим
такие $l$ и $r$, что
\[
 \textstyle
 \langle\,\Omega_{(-\infty,l)}\,\rangle
 \,=\,
 \langle\,\Omega_{(r,+\infty)}\,\rangle
 \,=\,
 \frac12-\frac12\sqrt{1-\cos^21} \,\approx\,
 0,07926\dots
\]
Легко видеть, что $l$ и $r$ существуют\footnote{Но, вообще говоря,
не единственны. Чтобы устранить эту многозначность, удобно
выбирать $l$ максимальным из возможных, а $r$~--- минимальным.
Тогда расстояние $r-l$ будет минимальным.}. При этом величину
$\delta_\rangle X = r-l$ вполне естественно можно назвать \rql
неопределённостью\rqr{} координаты $X$.

\mytheorem{}
 {\it Имеет место неравенство (принцип неопределённости):
\begin{equation}
 \begin{array}{|ccc|}
 \hline
 \phantom{1} & & \phantom{1}\\
  & \delta_\rangle X \Delta_\rangle P \geqslant \hbar & \\
 \phantom{1} & & \phantom{1}\\
 \hline
 \end{array}
 \label{Certainty}
\end{equation}}

Для доказательства этой теоремы оценим вначале скалярное
произведение вектора $\rangle$ и сдвинутого вектора
$e^{-\,i\,\delta_\rangle X\,P/\hbar} \,\rangle$:
\[
 \big|\,
  \langle\,
   |\,
   e^{-\,i\,\delta_\rangle X\,P/\hbar}
  \,\rangle
 \,\big|
 \,=\,
 \big|\,
  \langle\,
   (\,\Omega_{(-\infty,r)} + \Omega_{(r,+\infty)}\,)
   \,
   e^{-\,i\,\delta_\rangle X\,P/\hbar}
  \,\rangle
 \,\big|
 \,=\,
\]
\[
 \,=\,
 \big|\,
  \langle\,
   \,\Omega_{(-\infty,r)}\, e^{-\,i\,\delta_\rangle X\,P/\hbar}
  \,\rangle
   \,+\,
  \langle\,
   \Omega_{(r,+\infty)}\, e^{-\,i\,\delta_\rangle X\,P/\hbar}
  \,\rangle
 \,\big|
 \,\leqslant\,
\]
\[
 \,\leqslant\,
 \big|\,
  \langle\,
   \,\Omega_{(-\infty,r)}\, e^{-\,i\,\delta_\rangle X\,P/\hbar}
  \,\rangle
 \,\big|
   \,+\,
 \big|\,
  \langle\,
   \Omega_{(r,+\infty)}\, e^{-\,i\,\delta_\rangle X\,P/\hbar}
  \,\rangle
 \,\big|
 \,=\,
\]
\[
 \,=\,
 \big|\,
  \langle\,
   \,(\Omega_{(-\infty,r)})^2\, e^{-\,i\,\delta_\rangle X\,P/\hbar}
  \,\rangle
 \,\big|
   \,+\,
 \big|\,
  \langle\,
   (\Omega_{(r,+\infty)})^2\, e^{-\,i\,\delta_\rangle X\,P/\hbar}
  \,\rangle
 \,\big|
 \,=\,
\]
\[
 \,=\,
 \big|\,
  \langle\,
   \,\Omega_{(-\infty,r)}\,
   e^{-\,i\,\delta_\rangle X\,P/\hbar}
   \,\Omega_{(-\infty,l)}
  \,\rangle
 \,\big|
   \,+\,
 \big|\,
  \langle\,
   \Omega_{(r,+\infty)}\,
   e^{-\,i\,\delta_\rangle X\,P/\hbar}
   \,\Omega_{(l,+\infty)}
  \,\rangle
 \,\big|
 \,\leqslant\,
\]
\[
 \,\leqslant\,
  \langle\,
   \Omega_{(-\infty,r)}
  \,\rangle^{1/2}\,
  \langle\,
   \Omega_{(-\infty,l)}
  \,\rangle^{1/2}\,
   \,+\,
  \langle\,
   \Omega_{(r,+\infty)}
  \,\rangle^{1/2}\,
  \langle\,
   \Omega_{(l,+\infty)}
  \,\rangle^{1/2}\,
 \,=\,
\]
\[
 \textstyle
 \,=\,
 \sqrt{\frac12+\frac12\sqrt{1-\cos^21}}
 \cdot
 \sqrt{\frac12-\frac12\sqrt{1-\cos^21}}
 \ + \qquad\qquad\qquad\qquad
\]
\[
 \textstyle
 \qquad\qquad\qquad\qquad + \,
 \sqrt{\frac12-\frac12\sqrt{1-\cos^21}}
 \cdot
 \sqrt{\frac12+\frac12\sqrt{1-\cos^21}}
 \,=\,
 \cos 1
 \ .
\]
Здесь при переходе с пятой на шестую строку мы оценили оба
слагаемых с помощью неравенства Коши-Буняковского-Шварца.

Таким образом, для квантового угла между исходным и сдвинутым
векторами имеем оценку:
\[
 \angle\,\bigl(\
  \rangle
  \ , \
  e^{-\,i\,\delta_\rangle X\,P/\hbar}\,\rangle
 \ \bigr)
 \ \geqslant\ 1 \ .
\]
Но это означает, что под действием
 $e^{-\,i\,\delta_\rangle X\,P/\hbar}$
вектор $\rangle$ меняется {\it существенно}~\cite{Arb2005tcp}.

Применяя принцип {\it определённости}~\cite{Arb2005tcp},
непосредственно получаем~(\ref{Certainty}). $\blacksquare$


\paragraph{Обсуждение.}
Исторически сложилось так~\cite{HilgUff2001}, что сейчас во всех
учебниках принцип неопределённости Гейзенберга иллюстрируется с
помощью неравенства Кеннарда,
\[
 \Delta_\rangle X\,\Delta_\rangle P \,
 \geqslant\, \frac\hbar2 \ \ ,
\]
которое не является тождественным~(\ref{Certainty}). Его
несомненное достоинство состоит в том, что его доказательство
легче понять человеку, который впервые начинает изучать квантовую
механику. Однако, я всё же полагаю, что~(\ref{Certainty}) является
более фундаментальным.


\paragraph{Благодарности.}
В заключение хочу поблагодарить Т.~А.~Болохова, А.~В.~Осипова,
Е.~В.~Аксёнову,\\ А.~Ю.~Тощевикову, А.~Клейна, М.~Гатти,
А.~К.~Пэти и П.~Эндерса за полезные дискуссии, литературные
указания и помощь в добывании литературы.

